\documentclass[lettersize,journal]{IEEEtran}
\usepackage{amsmath,amsfonts,amssymb}
\usepackage{algorithm}
\usepackage{algpseudocode}
\usepackage{array}
\usepackage[caption=false,font=normalsize,labelfont=sf,textfont=sf]{subfig}
\usepackage{textcomp}
\usepackage{stfloats}
\usepackage{url}
\usepackage{verbatim}
\usepackage{graphicx}
\usepackage{multirow}
\usepackage{colortbl}
\usepackage[table]{xcolor}
\usepackage{arydshln}

\usepackage{booktabs}
\usepackage{diagbox}
\usepackage{cite}
\begin{document}

\title{EnCAgg: Enhanced Clustering Aggregation for Robust Federated Learning against Dynamic Model Poisoning}

\author{Tianyun Zhang, Zhen Yang, Haozhao Wang, Ru Zhang, Yongfeng Huang,~\IEEEmembership{Senior Member,~IEEE}
\thanks{This work was supported in part by the Natural Science Foundation of China under Grant U21B2020.(\textit{Corresponding author: Ru Zhang.})\par 
Tianyun Zhang, Zhen Yang contributes equally to this work. Tianyun Zhang, Zhen Yang and Ru Zhang are with School of Cyberspace Security, Beijing University of Posts and Telecommunications, Beijing 100876, China (email: zhangtianyun@bupt.edu.cn, yangzhenyz@bupt.edu.cn,  zhangru@bupt.edu.cn).\par
Haozhao Wang with the School of Computer Science and Technology, Huazhong University of Science and Technology, Wuhan 430074, China (e-mail: hz\_wang@hust.edu.cn).\par
Yongfeng Huang is with the Department of Electronic Engineering, Tsinghua University, Beijing 100084, China(email: yfhuang@tsinghua.edu.cn).}}

\markboth{Journal of \LaTeX\ Class Files,~Vol.~14, No.~8, August~2021}%
{Shell \MakeLowercase{\textit{et al.}}: A Sample Article Using IEEEtran.cls for IEEE Journals}

\IEEEpubid{0000--0000/00\$00.00~\copyright~2021 IEEE}

\maketitle

\begin{abstract}
Federated learning faces increasing threats from model poisoning attacks, which harms its application to improve privacy. Existing defense methods typically rely on fixed thresholds or perform clustering with a fixed number of clusters to distinguish malicious gradients from benign ones. However, these methods are difficult to adapt to dynamic poisoning strategies of malicious clients, and often result in the loss of benign gradients due to the heterogeneity of clients' local datasets. To address these problems, we propose a novel robust aggregation method that leverages a small number of known benign clients as references, enabling accurate identification and filtering of malicious gradients while retaining as many benign gradients as possible, even when the number of malicious clients is unknown and variable. First, we introduce a density-based low-dimensional gradient clustering method, which projects gradients onto the two most divergent dimensions and applies density-based clustering to identify malicious gradients while retaining clustered benign gradients and potentially benign outliers. Second, we design an enhancing clustering low-dimensional gradient generator model, which learns to generate pseudo-gradients aligned with the boundary of the benign cluster. These pseudo-gradients act as bridges to connect sparse benign gradient outliers. Third, we introduce low-dimensional gradient re-clustering that clusters the generated pseudo-gradients together with real gradients to recover benign gradients misclassified as noise points, enabling more benign gradients to participate in aggregation. Extensive experiments on the MNIST, CIFAR-10, and MIND datasets demonstrate that our method exhibits superior fidelity and robustness under dynamic poisoning scenarios.
\end{abstract}

\begin{IEEEkeywords}
federated learning, model poisoning attack, robust aggregation.
\end{IEEEkeywords}

\section{Introduction}
\IEEEPARstart{F}{ederated} learning (FL) allows decentralized clients to collaboratively train models without exposing raw data, and is widely used in privacy-sensitive domains. However, due to its distributed, heterogeneous and uncontrollable nature, FL is vulnerable to model poisoning attacks (MPAs), where malicious clients upload random or carefully crafted gradients to disrupt global convergence and degrade performance.

In particular, recent poisoning strategies have become increasingly stealthy and dynamic, substantially intensifying the challenge of defense methods. Sophisticated malicious clients can finely constrain their updates, minimizing distances or angular deviations from benign updates\cite{shejwalkar2021manipulating,10398506} or selectively perturb only high-impact parameters\cite{LI2024103936} to avoid detection. Besides, malicious clients may decide to alternate between benign and malicious behavior to fraudulently obtain a higher trust score, thereby making the malicious ratio vary across iterations. These stealthy and dynamic scenarios cause malicious updates to appear indistinguishable from benign ones in high-dimensional gradient space, especially since benign gradients naturally exhibit diversity.

Numerous robust aggregation methods have been proposed to defend MPAs. Traditional methods\cite{NIPS2017_f4b9ec30,pmlr-v80-yin18a,pmlr-v80-mhamdi18a,osti_10119268,sun2019reallybackdoorfederatedlearning} rely on fixed thresholds or statistical metrics to remove malicious gradients, usually assuming a known and relatively small proportion of adversaries. When the number of malicious clients fluctuates, fixed thresholds often lead to false positives or negatives. Moreover, due to data heterogeneity, the natural diversity among benign clients may cause some benign gradients to be mistakenly discarded, reducing generalization ability of the global model.

To flexibly defend against stealthy and dynamic attacks, recent studies have introduced clustering-based defenses. Some methods applied distance-based methods such as K-means, which partition all gradients into two clusters, assuming the larger one contains benign gradients\cite{10680604,10470437,11003999}. However, when no malicious gradients are present in an iteration, forced clustering may wrongly discard benign gradients. Besides, if malicious clients are dominant, they may form the largest cluster and evade detection. Other methods applied density-based methods like HDBSCAN, which divide gradients into multiple small clusters without requiring a preset number\cite{280048,ZHAO2025110990}. These approaches assign trust scores or weights based on features such as variance. However, benign updates may be inherently scattered and appear as outliers, while well-crafted malicious updates may form dense clusters that are mistakenly trusted as benign.

\IEEEpubidadjcol
In summary, current defense methods face two major challenges: first, most methods assume a fixed and relatively low proportion of malicious updates, but in reality, the behavior of malicious clients may vary dynamically across iterations. Some iterations may have no active attacks, while others may involve all malicious clients submitting poisoned gradients, resulting in a high proportion of adversarial gradients that disrupt detection accuracy. Second, benign gradients naturally exhibit diversity due to variations in client data, and traditional filtering methods often discard a considerable amount of useful updates. Nevertheless, we observe that the discrepancy between malicious and benign gradients is generally greater than the variation within benign gradients, while malicious gradients themselves tend to be more homogeneous—this offers a potential clue for precise discrimination.

To address the above issues, we propose a novel robust aggregation method for FL that does not rely on prior knowledge of the malicious client proportion or access to trusted data. By leveraging a small number of known benign clients as references, our method can accurately distinguish and filter out malicious gradients even under scenarios involving dynamic poisoning strategies from malicious clients. Specifically, we analyze the dissimilarity between client gradients and project high-dimensional gradients into a low-dimensional space to enhance the separability of benign and malicious gradients. We then apply density-based clustering in this low-dimensional space to identify potential benign clusters and introduce a generative model to aid in identifying borderline benign gradients, thereby improving the retention of sparse benign gradients. Finally, we employ a re-clustering mechanism to make more robust aggregation decisions, significantly enhancing resilience against dynamic poisoning attacks.

Our main contributions are summarized as follows: \par
1) We propose a density-based low-dimensional gradient clustering method that projects high-dimensional gradients onto a two-dimensional space using the two most divergent dimensions, and performs initial density-based clustering. This method effectively distinguishes benign from malicious gradients and retains possible benign gradients, overcoming the limitations of traditional fixed-threshold strategies.\par

2) We design an enhancing clustering low-dimensional gradient generator model. It is trained to generate pseudo-gradients along the boundary of the benign cluster, serving as “bridges” to connect benign gradients that were misclassified as outliers in the first round of clustering.  Then we introduce low-dimensional gradient re-clustering, which combines real and generated pseudo-gradients for a second density-based clustering to recover benign outliers. After re-clustering, the generation model is optimized using the clustering labels of the generated gradients along with their symmetry and diversity as constraints.\par

3) We conduct extensive experiments on three representative datasets(MNIST, CIFAR10 and MIND) to comprehensively evaluate the fidelity and robustness of our proposed method. The results demonstrate that our method consistently outperforms baseline methods under both low and high malicious ratios and effectively maintains superior global model performance.

\section{Related Work}
In this section, we review existing works on model poisoning attacks and corresponding defenses in federated learning.

\subsection{Model Poisoning Attacks}
Model poisoning attacks aim to craft or manipulate client-submitted gradients to introduce imperceptible interference during aggregation, thereby degrading global model performance and disrupting federated learning. The aim is to significantly degrade the performance of the global model and disrupt the normal operation of the federated system.

Early strategies manipulated all parameters using extreme values. For example, the Gaussian attack \cite{247652} samples malicious updates from a Gaussian distribution estimated from benign ones. Li et al. \cite{Li_Xu_Chen_Giannakis_Ling_2019} proposed (1) gradient explosion—uploading overly large gradients to cause instability, and (2) sign-flipping—reversing the sign of benign gradients to induce inverse learning. Tolpegin et al. \cite{10.1007/978-3-030-58951-6_24} manipulate local training data via label flipping and malicious sample injection to indirectly control the gradients uploaded by compromised clients. Wang et al. \cite{NEURIPS2020_b8ffa41d} proposed the model replacement attack, where a malicious client scales its update to dominate aggregation and effectively overwrite the global model.

However, overly aggressive perturbations are easily detected and removed by robust aggregation rules. To evade robust defenses, later methods tuned poisoning intensity. Baruch et al. \cite{NEURIPS2019_ec1c5914} used the negative mean of benign updates, allowing a few attackers to be effective. Fang attack \cite{247652} optimized perturbation strength coordinate-wise to bypass outlier detection like Median. UA-FedRec \cite{10.1145/3580305.3599923} extended such techniques to federated news recommendation, targeting news vectors, user embeddings, and sample sizes. FedGhost \cite{10877716} simulated local training trajectories to craft adversarial updates without real data. These approaches required manual tuning and showed unstable performance across tasks.

To achieve effective yet stealthy attacks, recent methods aim to disguise malicious updates as benign in terms of key statistical indicators such as mean and direction. Min-Max and Min-Sum attack  \cite{shejwalkar2021manipulating} employs multi-objective optimization balancing attack strength and detectability. Covert attack \cite{10121613} constrains distance, direction, and norm to make malicious updates statistically indistinguishable from benign ones.

Another direction selectively poisons key parameters. FedIMP \cite{LI2024103936} perturbs only high-impact parameters measured via Fisher information, achieving significant attack performance while keeping most parameters unchanged. SINE attack \cite{10398506} exploits cosine similarity by aligning perturbations with benign directions to boost evasion. DMPA \cite{feng2025dmpamodelpoisoningattacks} leverages model diversity across decentralized settings, introducing divergent perturbations across federated subgroups to escape clustering-based defenses.

Overall, poisoning strategies increasingly emphasize stealth and precision—injecting small, carefully crafted perturbations that degrade the model while eluding detection, posing growing challenges to existing defenses.

\subsection{Robust Aggregation Defenses}
The basic aggregation method, FedAvg/FedSGD \cite{pmlr-v54-mcmahan17a}, computes a weighted average of client updates based on local data size. To counter increasingly sophisticated poisoning attacks, robust aggregation methods have emerged, generally categorized as: (1) value/distance-based, (2) direction-based, (3) clustering-based and (4) reference-based defenses.

\subsubsection{Value/Distance-based Defenses}
These approaches use gradient magnitudes or pairwise distances to detect outliers or select representative values for aggregation. Krum \cite{NIPS2017_f4b9ec30} selects the update with the smallest total distance to others as the global update. Median and Trimmed-Mean \cite{pmlr-v80-yin18a} apply value selection, Median chooses the middle value, while Trimmed-Mean removes extremes before averaging. Bulyan \cite{pmlr-v80-mhamdi18a} enhances Krum with Trimmed-Mean. FABA \cite{osti_10119268} filters updates far from the mean. Norm-Bounding \cite{sun2019reallybackdoorfederatedlearning} constrains update norms. While these methods are simple and efficient, they may exclude a large number of clients, resulting in less representative global models.

\subsubsection{Direction-based Defenses}
These defenses assume benign gradients share directionality, while malicious ones deviate. FLTrust \cite{DBLP:conf/ndss/CaoF0G21} computes cosine similarity with a trusted gradient from a server-held root dataset. ShieldFL \cite{DBLP:journals/tifs/MaMMLD22} and TDFL \cite{9887909} replace server references with inter-client similarities. FLAIR \cite{10.1145/3579856.3582836} detects directional flips, and FLShield \cite{10646613} combines validation accuracy with cosine similarity. Although effective against extreme perturbations, these methods can be fooled by attacks which mimic the direction of benign updates. Moreover, the natural direction diversity among benign clients further reduces accuracy.

\subsubsection{Clustering-based Defenses}
These methods utilize various clustering techniques to aid in judgment, based on the assumption that benign updates exhibit strong consistency in direction and distance, while malicious updates cluster abnormally in these dimensions.

MODEL \cite{10680604} and DPFLA \cite{10470437} apply K-means to divide updates into two clusters, treating the larger one as benign. FLGuardian \cite{11003999} clusters per-layer metrics to detect layer-specific poisoning. However, these methods fail if attackers outnumber benign clients. FLAME \cite{280048} introduces random grouping followed by group-wise HDBSCAN clustering to detect local anomalies within each group. FedMP \cite{ZHAO2025110990} applies HDBSCAN globally and selects representative updates from each cluster for aggregation. However, HDBSCAN may over-segment benign updates, and lacking a reference, the server struggles to identify benign clusters.

\subsubsection{Reference-based Defenses}
To improve detection accuracy by establishing a reliable baseline, Some methods use clean server-side datasets to establish baselines. FLTrust \cite{DBLP:conf/ndss/CaoF0G21} generates a trusted gradient for directional filtering. FLShield \cite{10646613} uses clean validation sets. Zeno \cite{pmlr-v97-xie19b} compares loss on auxiliary data to detect adversaries. But these methods face severe performance degradation in dynamic model poisoning scenarios.

\section{Problem Statement}
In this section, we introduce our system model, describe the threat model, and define our defense objectives.

\begin{figure}[ht]
  \centering
  \includegraphics[width=1.0\linewidth]{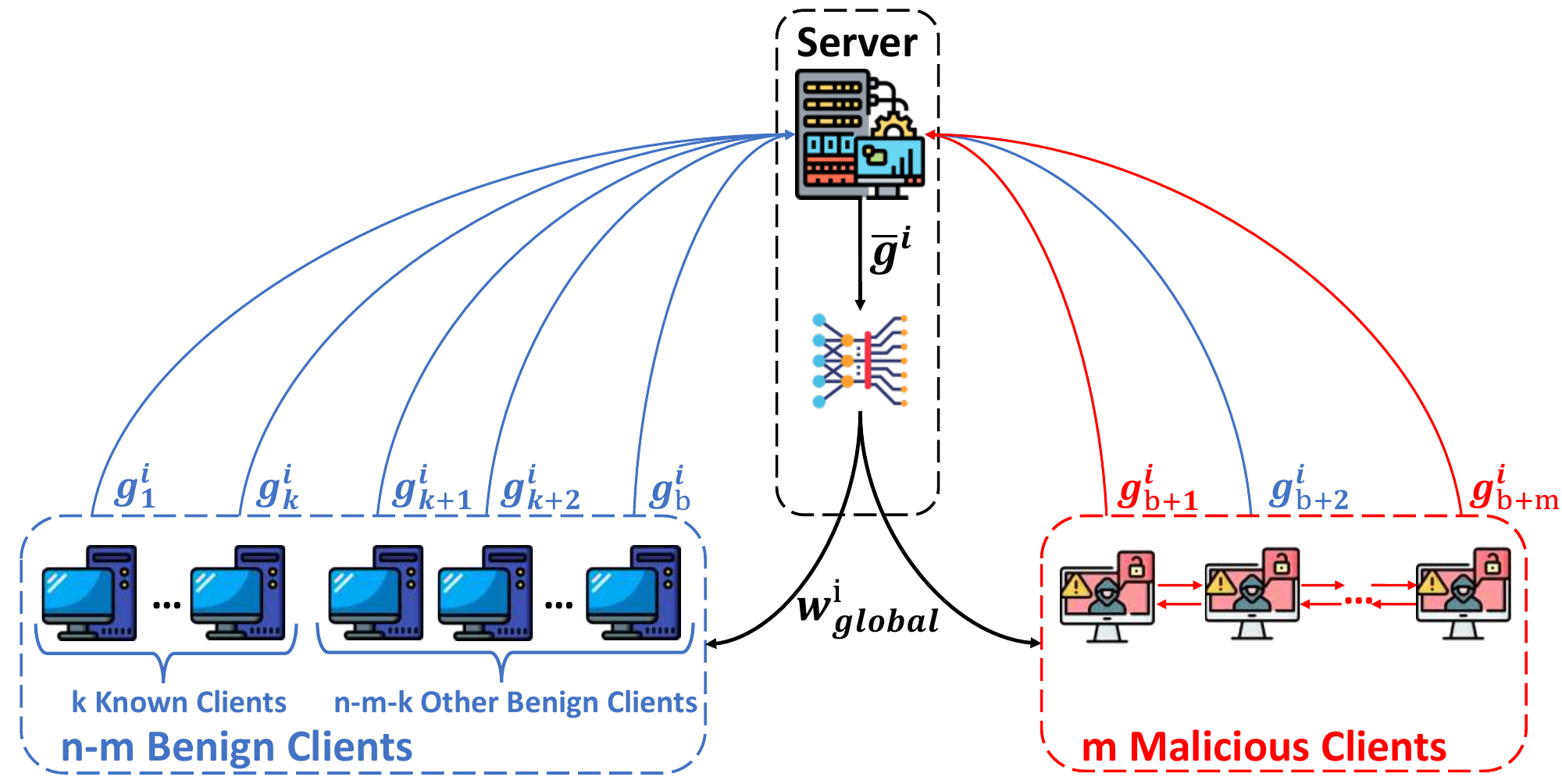}
  \caption{System Model}
  \label{fig:system_model}
\end{figure}

\subsection{System Model}

As illustrated in Fig.~\ref{fig:system_model}, we consider a federated learning system with $n = b + m$ clients, where $b$ are benign and $m$ are malicious. A central server coordinates training across $I$ iterations. In each iteration $i$, clients perform local training and upload gradients, while the server aggregates them to update the global model.

\begin{itemize}

    \item Clients: Each client holds a private dataset and a copy of the global model from the previous round. In iteration $i$, each benign client trains locally and uploads local gradients. Malicious clients, in contrast, craft adversarial gradients to degrade global performance. These attackers may communicate and share gradients, but they make independent per-iteration decisions on whether to poison or submit genuine gradients to gain trust. All clients upload gradients $g$ and await the updated global model.
    
    \item Server: Without access to client data or knowledge of the exact number of poisoned gradients per round, the server performs gradient screening. It knows $k$ clients ($2 \leq k \leq \frac{b}{2}$) are strictly benign, and uses their gradients as references to filter others. The filtered gradients are aggregated into a global update $\bar{g}$, which is used to update the model parameters $w$ before being sent to all clients.

\end{itemize}

\subsection{Threat Model}
We focus on model poisoning attacks (MPA), where malicious clients alter submitted gradients to disrupt global learning, without modifying local datasets. We assume each attacker can access the global model and its local dataset, and may obtain benign gradients from local data to guide attack generation. Poisoned updates may be shared across attackers, but the aggregation strategy remains unknown to them.

Furthermore, each malicious client adopts a dynamic strategy, which they decide independently in each round whether to poison or behave benignly.

\subsection{Defense Objectives}
We aim to design a robust aggregation method that dynamically handles varying numbers and behaviors of malicious clients while retaining as many benign contributions as possible. Specifically, our method should ensure:

\begin{itemize}
\item Robustness: The method should effectively filter diverse and dynamic attacks, eliminating only malicious gradients while retaining benign ones even when some clients alternate between malicious and benign behavior.

\item Fidelity: The method should minimize performance loss. When attackers are present, it may tolerate mild degradation due to reduced data, but must prevent significant global model corruption.
\end{itemize}

\begin{figure*}[ht]
  \centering
  \includegraphics[width=1.0\linewidth]{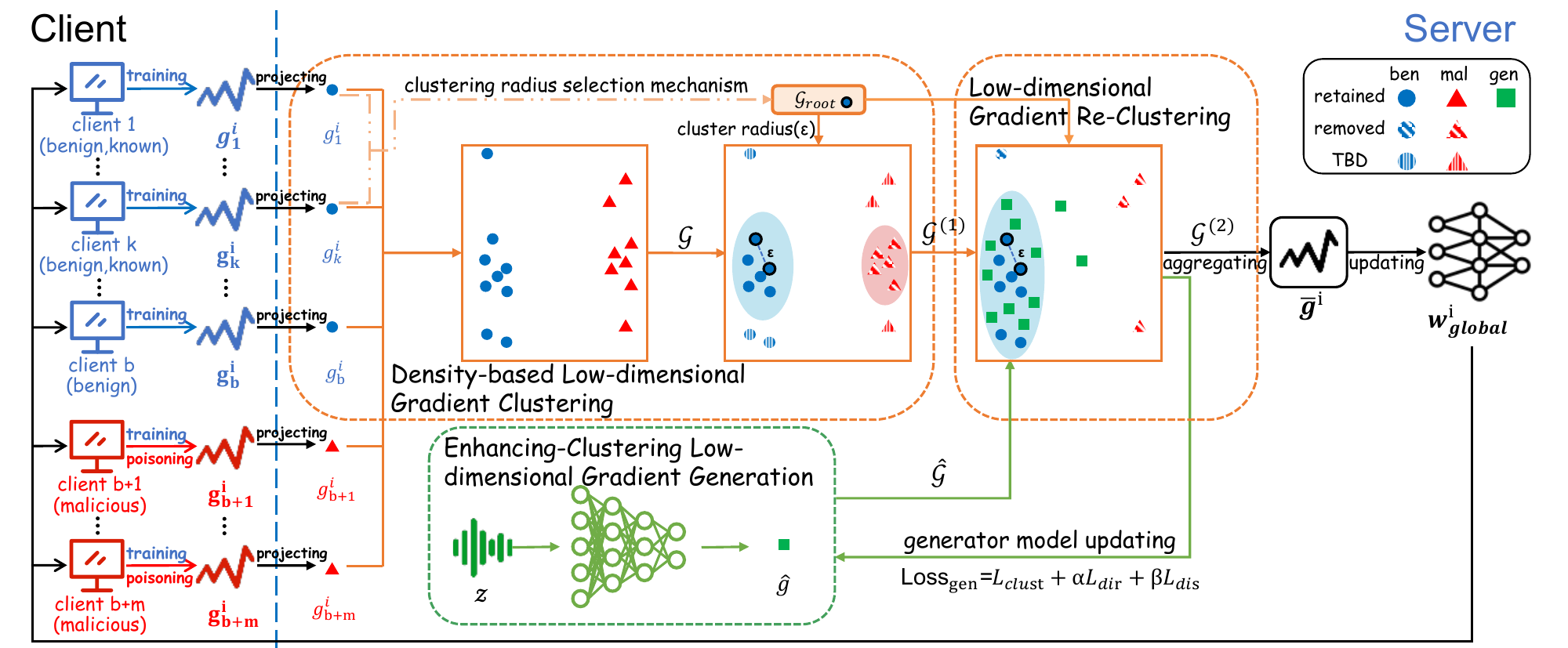}
  \caption{Framework of our method}
  \label{fig:framework}
\end{figure*}
 
\section{Our Method}
This section elaborates on the design of the method.
\subsection{Overview}
Fig. \ref{fig:framework}. shows the framework of our method, which consists of three main parts:
\begin{itemize}
    \item Distance-based Low-dimensional Gradient Clustering: The server computes pairwise differences among gradients submitted by clients to identify the two most variant dimensions, projecting gradients into a low-dimensional space to enhance separability between benign and malicious gradients. Density-based clustering then splits low-dimensional gradients into clusters, while server applies filter algorithm to exclude possible malicious gradients and retain others for next steps.

    \item Enhancing-Clustering Low-dimensional Gradient Generation: A generator model produces pseudo-gradients near the benign cluster boundary, linking it to sparsely located benign gradients previously misclassified as noise, thereby improving their likelihood of being correctly clustered in the next stage.

    \item Low-dimensional Gradient Re-Clustering: Retained real gradients and pseudo-gradients are combined for a second clustering. Sparse benign gradients can now join the benign cluster via the bridges built by pseudo-gradient, while distant noise points, which are more likely malicious now, are excluded. Only high-dimensional gradients corresponding to the real gradients in final benign cluster are aggregated, while pseudo-gradients are excluded.
\end{itemize}

\subsection{Density-based Low-dimensional Gradient Clustering}
Clustering gradients in high-dimensional space suffers from sparsity and large intra-class variance, leading to benign gradients being misclassified as noise and reducing effective samples. Some attacks further exploit this by perturbing only a few dimensions to mimic benign gradients. Moreover, methods like k-means require a fixed cluster number, which is unsuitable in dynamic attack scenarios which either excluding benign updates when no attack exists or misidentifying malicious clusters as benign under severe attacks.

To address this, we apply dimensionality reduction to enhance gradient separability and alleviate sparsity. Then, density-based clustering like DBSCAN adaptively detects clusters without predefined numbers, preserving benign gradients in normal cases and isolating outliers under attack. With a few trusted benign references, the benign cluster can be identified more reliably, enabling adaptive and precise anomaly detection in federated learning.

\begin{figure}[ht]
  \centering
  \includegraphics[width=1.0\linewidth]{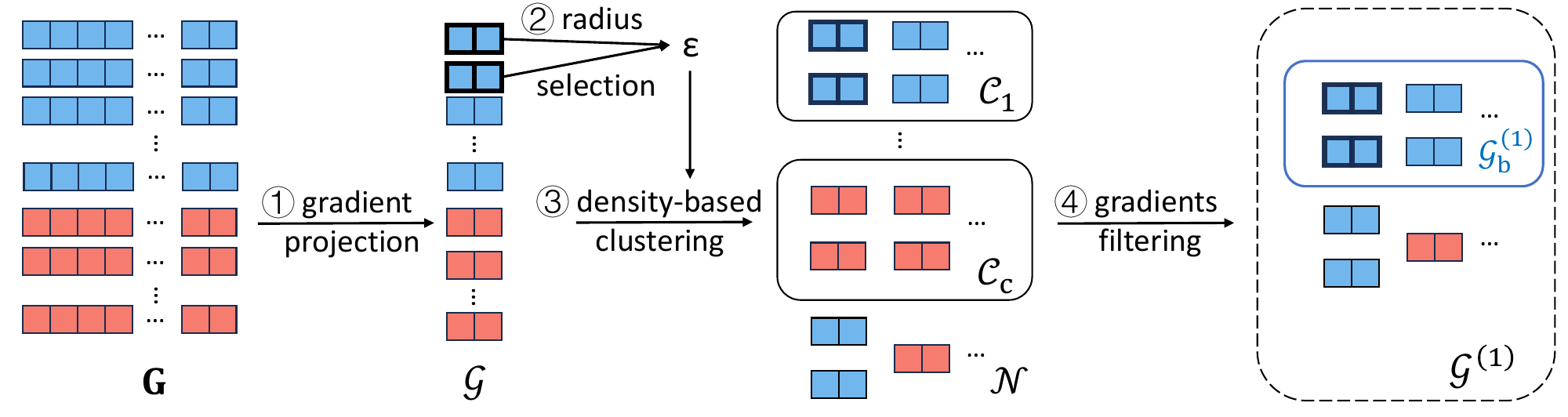}
  \caption{Density-based Low-dimensional Gradient Clustering}
  \label{fig:clustering}
\end{figure}

Based on the above analysis, we design a preliminary screening mechanism that combines gradient projection and density-based clustering. As shown in Fig. \ref{fig:clustering}., the process consists of four steps. First, the server analyzes all submitted gradients and identifies the two dimensions with the greatest variance. The original high-dimensional gradients are projected into a two-dimensional space. Then, the server selects two clients from the $k$ known benign clients and uses their distances to define a reasonable clustering radius. A density-based clustering is applied to the low-dimensional gradients, resulting in a clusters set $\mathcal{C}$ of size $c$ with corresponding cluster centers set $\mathbf{C}$ and remaining noise $\mathcal{N}$. Next, the server selects a benign cluster $ \mathcal{C}_b $ and retains gradients $ \mathcal{G}^{(1)}_b $ according to a predefined rule, along with the noise points $ \mathcal{N} $. These together form the retained gradient set $ \mathcal{G}^{(1)} $ after the first clustering.

\subsubsection{Gradient Projecting}

To address the challenge of detecting malicious gradients in high-dimensional space, we propose using Principal Component Analysis (PCA) to analyze the original high-dimensional gradients and project them into a lower-dimensional space by selecting the two dimensions with the greatest variance among gradients. Specifically, let the original gradient matrix be $\mathbf{G} \in \mathbb{R}^{n \times d}$ (where each high-dimensional gradient is denoted as $\vec{g}_i \in \mathbb{R}^{d}$ for $i = 1, 2, \ldots, n$). We compute the covariance matrix of $\mathbf{G}$ and obtain its top two eigenvectors corresponding to the largest eigenvalues. Let $V_2 \in \mathbb{R}^{d \times 2}$ denote the matrix composed of these two eigenvectors (principal directions), then the projected two-dimensional gradients are given by:
\begin{equation}
\mathcal{G} = \mathbf{G} V_2, \quad \mathcal{G} \in \mathbb{R}^{n \times 2}
\end{equation}
where each projected gradient is $g_i \in \mathbb{R}^2$ for $i = 1, 2, \ldots, n$.

Compared to direct detection in the original high-dimensional space, PCA projection maximizes the variance in the principal directions, thereby amplifying the anomalous deviations introduced by malicious gradients. In other words, if malicious gradients differ significantly from benign ones in certain dimensions, those dimensions tend to dominate the principal components, increasing the Euclidean distance $\|g_i - g_j\|_2$ between malicious and benign after projection.

This can be further interpreted as follows: suppose the difference vector between a malicious and benign gradient in the original space is $\Delta \vec{g}$, then the projected difference becomes $V_2^\top \Delta \vec{g}$. Its magnitude is controlled by the eigenvalues associated with the principal directions. Since principal components correspond to directions with the highest variance, and malicious perturbations typically induce significant variation in at least one direction, these components dominate the projection, thereby stretching the distance between benign and malicious updates in the projected space.

By projecting all gradients $\mathbf{G}$ into a 2D feature space $\mathcal{G}$ using PCA, we enhance the separability between benign and malicious gradients, providing a clearer and more discriminative feature representation for subsequent clustering process.

\subsubsection{Clustering Radius Selection}
In density-based clustering, the radius parameter $\varepsilon$ plays a crucial role. If it is set too small or too large, clustering quality may deteriorate. In federated training, the magnitude and direction of gradients change dynamically with the model state, especially in early rounds before convergence. Thus, a fixed radius across all rounds is infeasible: a small $\varepsilon$ may cause even similar benign gradients to be misclassified as noise, while a large $\varepsilon$ may mistakenly group distant malicious gradients into the benign cluster.

Given that the server knows $k$ strictly benign clients, we design a clustering radius selection mechanism. Specifically, the server randomly chooses two clients from the known benign set and uses the Euclidean distance between their projected gradients as the clustering radius. However, blindly selecting any two clients may introduce bias: if the distance is too small, the cluster will not cover most benign updates; if too large, it may admit malicious gradients.

Let the known benign gradient set be 
\begin{equation}
    \mathcal{G}_{\text{known}} = \{g_1, g_2, \dots, g_k\}.
\end{equation}
We compute all pairwise distances:
\begin{equation}
    d_{ij} = \|g_i - g_j\|_2, \quad \text{where } i, j \in [1, k],\ i < j.
\end{equation}
this yields $\binom{k}{2}$ distances forming a set $\mathcal{D}$, which is then sorted in ascending order:
\begin{equation}
    \mathcal{D}_{\text{sorted}} = \{ d_{(1)} \leq d_{(2)} \leq \dots \leq d_{(\binom{k}{2})} \}.
\end{equation}
The server selects the $\lceil r \cdot \binom{k}{2} \rceil$-th smallest value as the radius:
\begin{equation}
\varepsilon = d_{(\lceil r \cdot \binom{k}{2} \rceil)},
\end{equation}
where $r \in (0, 1)$ is a preset radius coefficient that determines which two known benign clients are selected as root clients $\mathcal{G}_{\text{root}}$. The Euclidean distance between their projected gradients is then used as the clustering radius $\varepsilon$. In practice, our primary goal is to prevent stealthy malicious gradients from being clustered into the benign group, while also aiming to include as many benign gradients as possible. Since most benign gradients are tightly clustered with small pairwise distances, whereas malicious ones lie farther away. Thus, choosing a moderately small $r$ yields a clustering radius $\varepsilon$ that effectively excludes malicious gradients while still covering the majority of benign updates. Besides, malicious gradients are typically closer to each other than to benign ones so that they are likely to form a separate cluster. This conservative selection improves the quality of clustering and provides a secure basis for subsequent steps.

\subsubsection{Density-based Clustering}

We apply DBSCAN (Density-Based Spatial Clustering of Applications with Noise) to cluster the projected gradients. Unlike K-means, DBSCAN does not require specifying the number of clusters. Instead, it uses two parameters: the clustering radius $\varepsilon$ and the minimum number of neighbors $min\_samples$.

Based on density, DBSCAN divide all gradients $\{g_1, g_2, \dots, g_n\} \subset \mathbb{R}^2$ into a set of clusters $C$ a set of noise points $\mathcal{N}$:
\begin{equation}
\begin{gathered}
    \mathcal{C} = \{ \mathcal{C}_1, \mathcal{C}_2, \dots, \mathcal{C}_c \}, \quad \mathcal{C}_i \subset \mathbb{R}^2 \\
    \mathcal{N} = n - \sum_{i=1}^c |\mathcal{C}_i|  
\end{gathered}
\end{equation}
A group of points is assigned to a cluster if any one point has at least $min\_samples$ neighbors within radius $\varepsilon$. If a point has no neighbors within $\varepsilon$ , it is treated as a noise point. After clustering, server computes the mean of each cluster to obtain the cluster centers $ \mathbf{C}$. 

Benign gradients are expected to be more dispersed due to data heterogeneity, while malicious gradients—often coordinated—are tightly clustered. When there are enough malicious clients, their updates may form a distinct cluster. Conversely, if they are few, they may appear as isolated noise. In either case, DBSCAN can effectively separate them from benign updates under most attack scenarios.

\subsubsection{Gradients Filtering}

Since DBSCAN is an unsupervised clustering algorithm, each resulting cluster lacks an inherent label, making it impossible for the server to directly identify which cluster consists of benign gradients. Additionally, even known benign clients may be marked as noise due to their distance from other samples, preventing their gradients from being used as labeled references. This increases the difficulty of identifying the benign cluster.

To address this issue, we design a hierarchical cluster filtering mechanism guided by root clients in Algorithm \ref{alg:filter}.

\begin{algorithm}[htbp]
\caption{Gradients Filtering}
\label{alg:filter}

\hspace{1.5em}\textbf{Input:} Projected gradients $\mathcal{G}$, known benign set $\mathcal{G}_{\text{known}}$, root clients set $\mathcal{G}_{\text{root}}$, cluster sets $C$, noise set $\mathcal{N}$, cluster center sets $\mathcal{C}$, cluster radius $\varepsilon$, cluster minimum samples number $min\_samples$, density constraint $\gamma$

\hspace{1.5em}\textbf{Output:} Filtered set $\mathcal{G}^{(1)}$

\begin{algorithmic}[1]
\State Let $l_{\text{root}}$ be the label of $\mathcal{G}_{\text{root}}$
\If{$l_{\text{root}} \neq -1$}
    \State $l_{\text{benign}} \gets l_{\text{root}}$
\ElsIf{$c > 1$}
    \For{each cluster $C_j$ with center $\mathbf{C}_j$ and label $j \in \{0, \dots, c-1\}$}
        \State Compute $D_j \gets \tfrac{1}{2}\big(\|\mathbf{C}_j - g_{\text{root}_1}\|_2 + \|\mathbf{C}_j - g_{\text{root}_2}\|_2\big)$
    \EndFor
    \State $l_{\text{benign}} \gets \arg\min_i D_i$
\Else
    \State $l_{\text{benign}} \gets -1$
\EndIf
\State $\mathcal{C}_{\text{benign}} \gets \{g_i \in \mathcal{G} \mid l_i = l_{\text{benign}}\}$
\State $\mathcal{G}_b^{(1)} \gets \{g_i \in \mathcal{C}_{\text{benign}} \mid \exists g_k \in \mathcal{G}_{\text{known}}, \|g_i - g_k\|_2 \le \gamma \varepsilon\}$
\State $\mathcal{G}^{(1)} \gets \mathcal{G}_b^{(1)} \cup \mathcal{N}$
\State \Return $\mathcal{G}^{(1)}$
\end{algorithmic}
\end{algorithm}

When valid clusters exist, the server first checks the clustering labels $l_{\text{root}}$ of the two root benign users. Since the clustering radius is based on their distance, if they are clustered together, they must belong to the same cluster. If this label is not $-1$, the cluster to which they belong is directly regarded as the benign cluster:  
\begin{equation}
    l_{\text{benign}} = l_{\text{root}}, \mathcal{C}_{\text{benign}} = C_{l_{\text{root}}}\quad \text{if } l_{\text{root}} \neq -1.
\end{equation}

If root clients are marked as noise ($l_{\text{root}} = -1$) but more than one valid cluster exists ($n_{\text{clusters}} > 1$), the server computes the average euclidean distance between each cluster center in $\mathbf{C}$ and the two root gradients $g_{\text{root}}$. Due to significant distance differences between benign and malicious gradients, the cluster closest to the root clients is more likely to be benign. Let:
\begin{equation}
    c^* = \arg\min_{c \in \mathcal{C}} \frac{1}{2} \left( \| \mathbf{C}_c - g_{\text{root}_1} \|_2 + \| \mathbf{C}_c - g_{\text{root}_2} \|_2 \right),
\end{equation}
then assign:
\begin{equation}
    l_{\text{benign}} = c^*,  
    \mathcal{C}_{\text{benign}} = C_c^*
\end{equation}

In extreme cases, if root clients are not clustered and there is at most one cluster, $n_{\text{clusters}} \le 1$ and $l_{\text{root}} = -1$, the server cannot determine the benign cluster. It then computes the global mean of all gradients as a temporary center and assigns cluster label $-1$ to all gradients for further screening.

To defend against potential propagation attacks, we incorporate a filtering step controlled by density constraint parameter $\gamma$. In propagation attacks, malicious clients may collude to design a chain of stealthy adversarial gradients. These gradients form a sequence where each is within the clustering radius $\varepsilon$ of the next, allowing a malicious gradient that is close to the benign cluster to “drag” other malicious ones into the same cluster under DBSCAN, ultimately corrupting the aggregation. To mitigate this risk, we restrict final benign selections to those within a tighter radius $\gamma\varepsilon$ of known benign gradients. Specifically, the refined benign set $\mathcal{G}_b^{(1)}$ is defined as:
\begin{equation}
    \mathcal{G}_b^{(1)} = \left\{ g_i \mid l_i = l_{\text{benign}} \ \text{and} \ \exists g_k \in \mathcal{G}_{\text{known}}, \ \|g_i - g_k\|_2 \le \gamma\varepsilon \right\}
\end{equation}
Finally, we reserve gradients in $\mathcal{G}_b^{(1)}$ and noise set $\mathcal{N}$ for further detection.
\begin{equation}
    \mathcal{G}^{(1)} = \mathcal{G}_b^{(1)} \cup \mathcal{N} = \{g_1,g_2,...,g_{n_1}\}.
\end{equation}
This hierarchical filtering ensures robust identification of benign gradients under varying attack intensities: most malicious gradients are removed after the first clustering, and only gradients in the benign cluster or labeled as noise are retained. Notably, even if strong attacks cause some benign gradients to be marked as noise, the geometric optimization strategy still helps identify the most likely benign cluster, providing a reliable foundation for the next-stage processing.

\subsection{Enhancing-Clustering Low-dimensional Gradient Generation}

To ensure projection stability, we reapply PCA to the high-dimensional gradient matrix $\mathbf{G}^{(1)}$ corresponding to the retained gradients $\mathcal{G}^{(1)}$. Since PCA directions depend on dominant distributions,  the relative positions of benign and malicious gradients can shift unpredictably between rounds when benign and malicious ratio are similar, hindering consistent boundary learning. After filtering most malicious gradients, reapplying PCA yields more stable projections, helping the generator better capture the benign cluster’s geometry and generate boundary pseudo-gradients.

After the first clustering and filtering, the server retains gradients in the benign cluster and those labeled as noise. Noise points may include both malicious and benign gradients, but benign ones are generally closer to and more directionally aligned with the benign cluster. This allows intermediate points to act as bridges, enabling the clustering radius $\varepsilon$ to reach and absorb sparse benign gradients. We thus design a gradient generation method that produces multiple pseudo-gradients near the boundary of the benign cluster to connect sparse benign gradients and enhance clustering quality without participating in aggregation.

The pseudo-gradients are designed to stay within $\varepsilon$ of the benign cluster and align directionally, while also being within $\varepsilon$ of sparse benign gradients. This makes the latter reachable in DBSCAN and can be included in the benign cluster. 

The generative model takes as input a set of $n_{\text{gen}}$ noise vectors $\mathcal{Z} = {z_1, \dots, z_{n_{\text{gen}}}}$, each $z_i \in \mathbb{R}^{d_g}$. Each vector in $\mathcal{Z}$ is independently passed through a shared three-layer fully connected network as:

\begin{equation}
    h^{(\ell)} = \tanh(W^{(\ell)} h^{(\ell-1)} + b^{(\ell)}), \quad \ell = 1, 2, 3
\end{equation}

where $ W^{(\ell)} $ and $ b^{(\ell)} $ are the weight matrix and bias term of the $\ell$-th layer, respectively.

The output of the model consists of two components:

\begin{itemize}
  \item A 2-dimensional vector $ \hat{g} \in \mathbb{R}^2 $, representing the generated pseudo-gradient.
  \item A confidence score $ \hat{y} \in (0, 1) $, produced by a sigmoid activation, estimating the likelihood that $ \hat{g} $ belongs to the benign cluster:
\end{itemize}

\begin{equation}
    \hat{g} = \tanh(W_g h^{(3)} + b_g), \quad \hat{y} = \sigma(W_y h^{(3)} + b_y)
\end{equation}

where $ \sigma(\cdot) $ denotes the sigmoid function.

\subsection{Low-dimensional Gradient Re-Clustering}

The server merges $ n_{\text{gen}} $ pseudo-gradients $\hat{\mathcal{G}}$ and remaining gradients $\mathcal{G}^{(1)}$ to form the new input and conduct a second round of DBSCAN clustering, with the same parameters $(\varepsilon, min\_samples)$ as before.

Let the new gradients set be denoted by:
\begin{equation}
\begin{gathered}
    \hat{\mathcal{G}} = \{\hat{g}_1, \dots, \hat{g}_{n_{\text{gen}}}\},
    \mathcal{G}^{(1)} = \{g_1, \dots, g_\}.
\end{gathered}
\end{equation}

The second-round clustering follows the same procedure as the first, but only gradients within the identified benign cluster are retained. All others, including noise points and gradients in non-benign clusters, are discarded. The resulting real gradients are denoted as $\mathcal{G}b^{(2)}$. In rare cases where all root gradients are labeled as noise and only one cluster is found, the server falls back to using the known benign set $\mathbf{G}_{\text{known}}$ for aggregation. Pseudo-gradients $\hat{\mathcal{G}}$ are not included in the final aggregation and discarded after clustering.

Introducing multiple pseudo-gradients around the benign cluster boundary effectively expands its $\varepsilon$-neighborhood, shortening the distance to sparsely distributed benign gradients and increasing their inclusion likelihood. Since inter-class distances (benign vs. malicious) typically exceed intra-class variation, gradients still labeled as noise are likely adversarial and safely removed.

\subsection{Global Model Aggregation}

After the second round of clustering and filtering, the server retains the final set of real low-dimensional gradients $\mathcal{G}_b^{(2)}$. These gradients correspond to original high-dimensional gradients, which are directly used in the final aggregation step.

The server performs a unweighted average over these high-dimensional gradients to compute the global update:
\begin{equation}
    \bar{g} = \frac{1}{|\mathcal{G}_b^{(2)}|} \sum_{\vec{g}_i \in \mathcal{G}_b^{(2)}} \vec{g}_i
\end{equation}

We intentionally avoid using weighted averaging (e.g., based on client sample sizes) to prevent malicious clients from manipulating the weights by falsely reporting larger training sets.

Then  then global model is updated by:
\begin{equation}
    \mathbf{w} \leftarrow \mathbf{w} - \eta \cdot \bar{g}
\end{equation}
where $\eta$ denotes the learning rate. The updated global model $\mathbf{w}$ is distributed to all clients for the next training round.

\subsection{Generator Model Updating}
Simultaneously with the global model update, the server optimizes the generator model using clustering results from the current round. The objective is to enhance the generator’s ability to produce pseudo-gradients that facilitate the expansion of the benign cluster boundary.

\subsubsection{Clustering Classification Loss}
We encourage the generator to produce pseudo-gradients that are likely to be assigned to the benign cluster during re-clustering. Instead of using cluster labels (which may vary across rounds), we treat cluster assignment as a binary classification task: for each pseudo-gradient $\hat{g}_j$, the ground-truth label $y_j \in \{0, 1\}$ is $1$ if it was clustered into the benign cluster, and $0$ otherwise. The generator outputs a predicted probability $\hat{y}_j \in (0, 1)$, and we define a weighted binary cross-entropy loss:
\begin{equation}
    \mathcal{L}_{\text{clust}} = -\frac{1}{n_{\text{gen}}} \sum_{j=1}^{n_{\text{gen}}} \left[ w_1 y_j \log \hat{y}_j + w_0 (1 - y_j) \log (1 - \hat{y}_j) \right]
\end{equation}
Here, $w_1 > w_0$ to penalize false negatives more heavily and avoid trivial solutions where all generated samples are rejected, such as $y_j = 0, \hat{y}_j \approx 0$.

\subsubsection{Direction Diversity Loss}
To avoid generated gradients clustering tightly at the benign center, we encourage symmetric and dispersed directions. Let $\mu_x, \mu_y$ denote the mean offsets in the $x$ and $y$ directions of pseudo-gradients from the benign cluster center, and let $\sigma_x, \sigma_y$ be their respective standard deviations. We define:
\begin{equation}
\begin{gathered}
    \text{std}_x = \text{ReLU}(\tau - \sigma_x), \quad \text{std}_y = \text{ReLU}(\tau - \sigma_y) \\
    \mathcal{L}_{\text{dir}} = |\mu_x| + |\mu_y| + \text{std}_x + \text{std}_y
\end{gathered}
\end{equation}
where $\tau$ is a minimum dispersion threshold. This loss penalizes imbalance and collapse in directional spread.

\subsubsection{Distance Diversity Loss}
We further enforce spatial separation among generated pseudo-gradients. Define a target minimum distance $\rho \cdot \varepsilon$, where $\varepsilon$ is the clustering radius and $\rho$ is a scaling factor. For each pair of generated gradients $(\hat{g}_i, \hat{g}_j)$, we define the pairwise penalty:
\begin{equation}
    \mathcal{L}_{\text{dis}} = \frac{1}{n_{\text{gen}}} \sum_{1 \le i < j \le n_{\text{gen}}} \max(0, \rho \cdot \varepsilon - \|\hat{g}_i - \hat{g}_j\|_2)^2
\end{equation}
This ensures that pseudo-gradients are not overly concentrated.

\subsubsection{Total Generator Loss}
Combining the above, the total loss for the generator is:
\begin{equation}
    \mathcal{L}_G = \mathcal{L}_{\text{clust}} + \alpha \cdot \mathcal{L}_{\text{dir}} + \beta \cdot \mathcal{L}_{\text{dis}}
\end{equation}
where $\alpha$ and $\beta$ are hyper-parameters controlling the trade-off between classification accuracy and diversity.

\section{Theoretical analysis}
In this section, we conduct the necessary theoretical analysis. We separately analyze the convergence of the generator model and the global model.

\subsection{Convergence Analysis of Generator Model}

We analyze the convergence of the generative model in EncAgg. The following assumptions are made:

\textbf{Assumption 1:} The objective function $\mathcal{F}_G(\theta)$ is $L$-smooth, that is, its gradient is $L$-Lipschitz continuous:
\begin{equation}
\|\nabla \mathcal{F}_G(\theta_1) - \nabla \mathcal{F}_G(\theta_2)\| \le L \|\theta_1 - \theta_2\|,\quad \forall \theta_1,\theta_2.
\end{equation}

\textbf{Assumption 2:} $\mathcal{F}_G(\theta)$ has a non-negative lower bound: $\mathcal{F}_G(\theta) \ge \mathcal{F}_{\min} \ge 0$.

\textbf{Assumption 3:} The generator uses a fixed learning rate $\eta \le \frac{1}{L}$ and performs single-step gradient descent:
\begin{equation}
\theta_{t+1} = \theta_t - \eta \nabla \mathcal{F}_G(\theta_t)
\end{equation}

\textbf{Lemma 1:}  
Under Assumptions 1–3, we have the following descent inequality:
\begin{equation}
\mathcal{F}_G(\theta_{t+1}) \le \mathcal{F}_G(\theta_t) - \left( \eta - \frac{L \eta^2}{2} \right) \| \nabla \mathcal{F}_G(\theta_t) \|^2
\end{equation}
In particular, when $\eta \le \frac{1}{L}$:
\begin{equation}
\mathcal{F}_G(\theta_{t+1}) \le \mathcal{F}_G(\theta_t) - \frac{\eta}{2} \| \nabla \mathcal{F}_G(\theta_t) \|^2
\end{equation}

\textbf{Proof:}  
Since $\mathcal{F}_G$ is $L$-smooth, that means we can apply the standard upper bound of the first-order Taylor expansion for smooth non-convex functions:
\begin{equation}
\mathcal{F}_G(\theta_{t+1}) \le \mathcal{F}_G(\theta_t) + \nabla \mathcal{F}_G(\theta_t)^\top (\theta_{t+1} - \theta_t) + \frac{L}{2} \|\theta_{t+1} - \theta_t\|^2
\end{equation}
Substituting the update rule $\theta_{t+1} = \theta_t - \eta \nabla \mathcal{F}_G(\theta_t)$ into the inequality yields:
\begin{equation}
\mathcal{F}_G(\theta_{t+1}) \le \mathcal{F}_G(\theta_t) - \eta \| \nabla \mathcal{F}_G(\theta_t) \|^2 + \frac{L}{2} \eta^2 \| \nabla \mathcal{F}_G(\theta_t) \|^2
\end{equation}
which simplifies to the desired descent inequality.

Summing over $T$ steps and using Assumption 2, we obtain:
\begin{equation}
\frac{1}{T} \sum_{t=0}^{T-1} \|\nabla \mathcal{F}_G(\theta_t)\|^2 \le \frac{2}{\eta T} \left( \mathcal{F}_G(\theta_0) - \mathcal{F}_{\min} \right)
\end{equation}
Since the average of $T$ non-negative values is bounded, there must exist at least one $t^* < T$ such that the gradient norm at $t^*$ is below the average. Therefore we can give Theorem 1.

\textbf{Theorem 1:}  
Under Assumptions 1–3, for any $\delta > 0$, if
\begin{equation}
T \ge \frac{2}{\eta \delta^2} \left( \mathcal{F}_G(\theta_0) - \mathcal{F}_{\min} \right),
\end{equation}
then there exists $t^* < T$ such that $\|\nabla \mathcal{F}_G(\theta_{t^*})\| \le \delta$.

\subsection{Convergence Analysis of Global Model}
To analyze the convergence of the global model with EnCAgg, the necessary assumptions are listed below:

\textbf{Assumption 1:} The non-convex loss function $F(w)$ is $L$-Lipschitz smooth, that is, for any $w, w'$, we have:
\begin{equation}
\|\nabla F(w) - \nabla F(w')\| \le L \|w - w'\|,
\end{equation}
where $\nabla F(w)$ is the gradient vector of $F(w) $about $w$.

\textbf{Assumption 2:} Under ideal conditions, the global model is obtained by averaging all benign clients. We assume the set of benign clients is $\mathcal{B}$.
For each benign client $i \in \mathcal{B}$, its local loss function is defined as $f_i(w)$. The global loss function is
\begin{equation}
F(w) = \frac{1}{|\mathcal{B}|} \sum_{i \in \mathcal{B}} f_i(w).
\end{equation}

We assume the client gradient differences are bounded in variance, and there exists a constant $\sigma \ge 0$ , such that:
\begin{equation}
\frac{1}{|\mathcal{B}|} \sum_{i \in \mathcal{B}} \|\nabla f_i(w) - \nabla F(w)\|^2 \le \sigma^2, \quad \forall w.
\end{equation}

\textbf{Assumption 3:} There exists a constant $v > 0$ such that for any benign client $ i \in \mathcal{B}$ and any $w$, we have:
\begin{equation}
\|\nabla f_i(w)\| \le v.
\end{equation}

Based on the above assumptions, we analyze the convergence of the global model under two scenarios: (1) high and (2) low-intensity malicious gradients.

\subsubsection{Case 1: High-Intensity Malicious Gradients}

When malicious gradients are strongly perturbed, their distances from benign gradients are large enough to be excluded during clustering. As a result, only benign gradients are aggregated, and the process reduces to the benign-only scenario.

By Assumption 2, the aggregated gradient equals the true gradient: $\mathbf{\tilde{g}}_t = \nabla F(w_t)$. Using the $L$-smoothness in Assumption 1 and the gradient descent update $w_{t+1} = w_t - \eta \nabla F(w_t)$, we obtain:
\begin{equation}
F(w_{t+1}) \le F(w_t) - \eta \|\nabla F(w_t)\|^2 + \frac{L}{2} \eta^2 \|\nabla F(w_t)\|^2.
\end{equation}

If $L\eta \le 1$, then:
\begin{equation}
F(w_{t+1}) \le F(w_t) - \frac{1}{2}\eta \|\nabla F(w_t)\|^2,
\end{equation}
showing that the loss decreases at each step as long as the gradient is non-zero. Thus, $F(w_t)$ is monotonically non-increasing and bounded below, implying the convergence.

Summing over $t = 0$ to $T - 1$ and using $F(w_T) \ge F_{\inf}$:
\begin{equation}
\frac{1}{T} \sum_{t=0}^{T-1} \|\nabla F(w_t)\|^2 \to 0.
\end{equation}
This implies that $\min_{0 \le t < T} \|\nabla F(w_t)\|^2 \to 0$, which means the global gradient converges to zero and $w_t$ approaches a stationary point.

\subsubsection{Case 2: Low-Intensity Malicious Gradients}

In this case, malicious gradients with small perturbations may be retained in the benign cluster with a big clustering radius, and may lead to a propagating attack, where malicious gradients are gradually absorbed into the benign cluster via radius-based chaining. However, under a density constraint parameter $\gamma$, we can still find a suitable $\varepsilon$ to prevent unrestricted malicious gradient propagation and ensure convergence.

\textbf{Assumption 4 (Special Condition for Case 2): } 
Let $\mathcal{G}_b^{(2)}$ denote the set of low-dimensional gradients retained after the second-round clustering, and let $g_m^{(2)} \subseteq \mathcal{G}_b^{(2)}$ denote the subset of malicious gradients that are mistakenly retained and participate in the final aggregation. We make the following additional assumption:

If DBSCAN clusters malicious gradients together with benign gradients, there must exist at least one benign gradient in low-dimensional space, denoted by $g_a$, such that:
\begin{equation}
\min_{g \in g_m^{(2)}} \|g - g_a\| < \varepsilon,
\end{equation}
Since we impose an upper bound $\gamma \varepsilon$ on the distance from each gradient in $\mathcal{G}_b^{(2)}$ to known benign gradients, we have:
\begin{equation}
\|g - g_a\| \le \lfloor\gamma\rfloor\varepsilon, \quad \forall g \in g_m^{(2)}.
\end{equation}

\textbf{Lemma 1:} If the distance between $g_m \in g_m^{(2)} \subseteq \mathcal{G}_b^{(2)}$ and a benign gradient $g_a \in \mathcal{G}_b^{(2)}$ satisfies:
\begin{equation}
\| g_m - g_a \| \le \gamma\varepsilon,
\end{equation}
Then their distance in the high-dimensional space satisfies:
\begin{equation}
\|\vec{g}_m - \vec{g}_a\|^2 \le \lfloor\gamma\rfloor^2 \varepsilon^2 + 2(n - 1)(d - 2)\lambda_2.
\end{equation}

\textbf{Proof:} Since PCA projects the gradients onto orthonormal directions $\{u_1, u_2, \ldots, u_d\}$, we can decompose the full high-dimensional squared distance as:
\begin{equation}
\begin{aligned}
\|\vec{g}_m - \vec{g}_a\|^2 &= \sum_{\ell=1}^{d} \left[(\vec{g}_m - \vec{g}_a)^\top u_\ell \right]^2 \\
&= \sum_{\ell=1}^{2} \left[(\vec{g}_m - \vec{g}_a)^\top u_\ell \right]^2 + \sum_{\ell=3}^{d} \left[(\vec{g}_m - \vec{g}_a)^\top u_\ell \right]^2
\end{aligned}
\end{equation}

The first term corresponds to the 2D projected distance:
\begin{equation}
\sum_{\ell=1}^{2} \left[(\vec{g}_m - \vec{g}_a)^\top u_\ell \right]^2 = \| (g_m - g_a) \|^2 \le \gamma^2 \varepsilon^2.
\end{equation}

To upper-bound the tail sum, consider an extreme case where only $\vec{g}_m$ and $\vec{g}_a$ lie at opposite ends along each direction $u_\ell$ ($\ell \ge 3$), and the remaining $n - 2$ samples lie at the mean. This maximizes the variance under the constraint $\lambda_\ell \le \lambda_2$:
\begin{equation}
\left[(\vec{g}_m - \vec{g}_a)^\top u_\ell \right]^2 \le 2(n - 1)\lambda_\ell \le 2(n - 1)\lambda_2.
\end{equation}

Summing over $\ell = 3$ to $d$ gives:
\begin{equation}
\sum_{\ell=3}^{d} \left[(\vec{g}_m - \vec{g}_a)^\top u_\ell \right]^2 \le 2(n - 1)(d - 2)\lambda_2.
\end{equation}

Hence, the full squared distance is bounded by:
\begin{equation}
\|\vec{g}_m - \vec{g}_a\|^2 \le \lfloor\gamma\rfloor^2 \varepsilon^2 + 2(n - 1)(d - 2)\lambda_2.
\end{equation}

This implies that each malicious gradient has an upper bound on its distance to the same benign gradient $g_a$. From this lemma, we can further discuss the convergence of global model. Let: 

\begin{equation}
    \kappa = s\lfloor\gamma\rfloor + u.
\end{equation}
Suppose $a=(a_1,...,a_j,...a_d)$ is opposite direction with the correct part of global gradient $\tilde{g}_t$, 
\begin{multline}
\tilde{g}_t \geq (n'-\kappa) \vec{g}_b^{(2)} + \kappa \vec{g}_a+s\sum_{i=1}^{\lfloor\gamma\rfloor}(a_1 \sqrt{\frac{i^2\varepsilon^2+2(n-1)(d-1)\lambda_2}{d}},\\...,a_d\sqrt{\frac{i^2\varepsilon^2+2(n-1)(d-1)\lambda_2}{d}})+\\ \sum_{i=1}^{u}(a_1 \sqrt{\frac{i^2\varepsilon^2+2(n-1)(d-1)\lambda_2}{d}},\\...,a_d\sqrt{\frac{i^2\varepsilon^2+2(n-1)(d-1)\lambda_2}{d}}),a_j=\pm1.
\end{multline}
Let 
\begin{equation}
\begin{aligned}
S_\gamma&=\sum_{i=1}^{\lfloor\gamma\rfloor}\sqrt{\frac{i^2\varepsilon^2+2(n-1)(d-1)\lambda_2}{d}},\\
S_u &= \sum_{i=1}^{u}\sqrt{\frac{i^2\varepsilon^2+2(n-1)(d-1)\lambda_2}{d}}
\end{aligned}
\end{equation}
Since
\begin{equation}
\resizebox{\linewidth}{!}{$
\begin{aligned}
S_\gamma&\leq\sum_{i=1}^{\lfloor\gamma\rfloor}(\frac{i\varepsilon}{\sqrt{d}}+\sqrt{2(n-1)\lambda_2})=\frac{(1+\lfloor\gamma\rfloor)\lfloor\gamma\rfloor\varepsilon}{2\sqrt{d}}+\lfloor\gamma\rfloor\sqrt{2(n-1)\lambda_2}, \\
S_u&\leq\sum_{i=1}^{u}(\frac{i\varepsilon}{\sqrt{d}}+\sqrt{2(n-1)\lambda_2})=\frac{(1+u)u\varepsilon}{2\sqrt{d}}+u\sqrt{2(n-1)\lambda_2}
\end{aligned}
$}
\end{equation}
Then,
\begin{multline}
s\cdot S_\gamma+S_u\leq
\frac{(s\lfloor\gamma\rfloor+s\lfloor\gamma\rfloor^2+u+u^2)\varepsilon}{2\sqrt{d}}\\+(s\lfloor\gamma\rfloor+u)\sqrt{2(n-1)\lambda_2}=\frac{(\kappa+\kappa\lfloor\gamma\rfloor+u^2-u\lfloor\gamma\rfloor)\varepsilon}{2\sqrt{d}}\\+\kappa\sqrt{2(n-1)\lambda_2}\leq\frac{(1+\lfloor\gamma\rfloor)\kappa\varepsilon}{2\sqrt{d}}+\kappa\sqrt{2(n-1)\lambda_2}
\end{multline}
Given $a=(a_1,...,a_j,...a_d)$ is opposite direction with the correct part of $\tilde{g}_t$, which we call $\vec{g}_{bb}$,
\begin{multline}
\vec{g}_{bb}=(\kappa+1)\vec{g}_a+\sum_{k=1}^{n'-\kappa}\vec{g}_{bk}=(g_{bb1},...,g_{bbj},...,g_{bbd}),\\
\tilde{g}_t \geq \vec{g}_{bb}+(\frac{(1+\lfloor\gamma\rfloor)\kappa\varepsilon}{2\sqrt{d}}+\kappa\sqrt{2(n-1)\lambda_2})(a_1,..,a_j,...,a_d)
\end{multline}

We need to satisfy this inequality to achieve convergence:
\begin{multline}
g_{bb1}(g_{bb1}+a_1(\frac{(1+\lfloor\gamma\rfloor)\kappa\varepsilon}{2\sqrt{d}}+\kappa\sqrt{2(n-1)\lambda_2})+...+\\g_{bbd}(g_{bbd}+a_d(\frac{(1+\lfloor\gamma\rfloor)\kappa\varepsilon}{2\sqrt{d}}+\kappa\sqrt{2(n-1)\lambda_2}))\geq 0
\end{multline}

Then we can derive:
\begin{equation}
\sum_{j=1}^{d}g_{bbj}^2-|g_{bbj}|(\frac{(1+\lfloor\gamma\rfloor)\kappa\varepsilon}{2\sqrt{d}}+\kappa\sqrt{2(n-1)\lambda_2})\geq 0
\end{equation}

It can be seen as equally:
\begin{equation}
\frac{\sum_{j=1}^{d}g_{bbj}^2}{\sum_{j=1}^{d}|g_{bbj}|}\geq\frac{(1+\lfloor\gamma\rfloor)\kappa\varepsilon}{2\sqrt{d}}+\kappa\sqrt{2(n-1)\lambda_2}
\end{equation}

To get an available $\varepsilon$, we need to let:
\begin{equation}
\varepsilon\leq
\frac{\sum_{j=1}^{d}g_{bbj}^2}{\sum_{j=1}^{d}|g_{bbj}|}\cdot\frac{2\sqrt{d}}{\kappa(1+\lfloor\gamma\rfloor)}-\frac{2\sqrt{2(n-1)d\lambda_2}}{1+\lfloor\gamma\rfloor}
\end{equation}

So a non-negative $\varepsilon$ requires:
\begin{equation}
(\frac{\sum_{j=1}^{d}g_{bbj}^2}{\sum_{j=1}^{d}|g_{bbj}|})^2\geq2\kappa^2(n-1)\lambda_2
\end{equation}

With the above inequality holding true, we can always find an appropriate $\varepsilon$ to guarantee global model convergence. This implies that convergence can be achieved under any proportion of malicious clients, as long as $\gamma$ is imposed and a non-negative $\varepsilon$ satisfying the inequality is chosen.

\section{Experiments}
In this section, we evaluate the performance of our model from multiple perspectives and compare it with existing works.

\subsection{Experimental Settings}
\subsubsection{Environment Setup}
Our method is implemented in Python 3.8. All experiments are conducted on a server equipped with Ubuntu 20.04, an Intel(R) Xeon(R) Gold 5220R CPU, 256 GB RAM, and an Nvidia GeForce RTX 3090 GPU. 

\subsubsection{Datasets and Metrics}
We evaluate the performance of our method on three datasets: MNIST, CIFAR-10, and MIND\cite{wu-etal-2020-mind}. Both MNIST and CIFAR10 are image classification datasets which have ten categories. MNIST has 60,000 training examples and 10,000
testing examples while CIFAR10 contains 50,000 training examples and 10,000 testing examples. On these datasets, we use model accuracy as evaluation metric.

MIND is a large-scale dataset built by Microsoft News Platform for news recommendation tasks. The dataset comes in two versions, MIND-small and MIND-large. In this work,we use MIND-small, which contains 65,238 news articles and 50,000 users. On MIND, we use AUC, MRR and NDCG as evaluation metrics. AUC measures model’s ability to recommend candidate news, while MRR and NDCG evaluate the ranking quality of the recommended results.

\begin{table}[htbp]
\centering
\caption{Federated learning configuration on different datasets}
\label{tab:fl-settings}
\resizebox{\linewidth}{!}{%
\begin{tabular}{l|l|c c c}
\toprule
\textbf{Parameter} & \textbf{Explanation} & \textbf{MNIST} & \textbf{CIFAR-10} & \textbf{MIND} \\
\midrule
model & Model & CNN & ResNet18 & NRMS \\
$n$ & Total clients & 20 & 20 & 50 \\
$E$ & Epochs & 10 & 100 & 5 \\
$I$ & Iterations per epoch & 50 & 50 & 1000 \\
$\eta$ & Learning rate & 0.001 & 0.01 & 0.0001 \\
$mal$ & Malicious Ratio & \multicolumn{3}{c}{10\% \& 60\%} \\
$min\_samples$ & Min cluster sample number & \multicolumn{3}{c}{5} \\
$r$ & Radius coefficient & \multicolumn{3}{c}{0.2} \\
$\gamma$ & Density Constraint & \multicolumn{3}{c}{3.0} \\
$n_{\text{gen}}$ & Pseudo Gradient Number & \multicolumn{3}{c}{100} \\
\bottomrule
\end{tabular}
}
\end{table}

\subsubsection{FL System Settings}
The FL system configurations are summarized in Table \ref{tab:fl-settings}. For MNIST, we use a CNN with two convolutional layers and one fully connected layer. For CIFAR-10, we adopt ResNet-18, which includes 17 convolutional layers and 1 fully connected layer with residual blocks and shortcut connections. For MIND, we employ NRMS \cite{wu-etal-2019-neural-news}, a model utilizing multi-head self-attention to represent user interests and match relevant news.

Each epoch consists of multiple iterations, during which each client trains on one local data batch, computes gradients, and uploads them to the server. The server then applies our EnCAgg algorithm to identify and filter potential malicious gradients. Remaining benign gradients are aggregated to update the global model. While we use dataset-specific settings for epoch count, iteration number, client count, and learning rate, we keep the radius coefficient $r$ and density constraint $\gamma$ consistent across datasets.

To evaluate robustness under varying adversarial intensities, we test at 10\% and 60\% malicious client ratios, which representing benign-dominant and adversary-dominant scenarios, respectively. Malicious clients follow a dynamic strategy, submitting either malicious or benign gradients per round, ensuring their gradient proportion never exceeds the predefined malicious ratio $mal$.

\subsubsection{Attack and Defense Methods}
In our experiments, the model is attacked using three model poisoning strategies:
\textbf{1. UA-FedRec}\cite{10.1145/3580305.3599923} is a targeted attack designed for news recommendation systems. It separately crafts the gradients of the user and news encoders to induce significant performance degradation using only a small fraction of malicious clients. On the MNIST and CIFAR-10 datasets, we adopt only the user encoder poisoning component to simulate this attack in general classification settings.
\textbf{2. LIE (Little is Enough)}\cite{NEURIPS2019_ec1c5914} perturbs benign gradients by adding small manipulations in the opposite direction of the benign mean. This allows malicious updates to evade detection methods based on thresholding or deviation metrics, while still degrading model performance.
\textbf{3. Min-Max}\cite{shejwalkar2021manipulating} seeks to maximize the adversarial effect of poisoned gradients while maintaining a bounded distance from benign updates. This design helps the attack bypass distance-based defense mechanisms by mimicking the distribution of legitimate gradients.

To evaluate the performance of our EnaAgg, we compare it with seven high-performing baselines, including FedSGD\cite{pmlr-v54-mcmahan17a}, Krum\cite{NIPS2017_f4b9ec30}, Median\cite{pmlr-v80-yin18a}, TrimmedMean\cite{pmlr-v80-yin18a}, FLTrust\cite{DBLP:conf/ndss/CaoF0G21}, DPFLA\cite{10470437} and FLGuardian\cite{11003999}.

\begin{table}[htbp]
\centering
\renewcommand{\arraystretch}{1.3}
\setlength{\tabcolsep}{5pt}
\caption{Fidelity Evaluation (\%)}
\label{tab:fidelity}
\resizebox{\linewidth}{!}{%
\begin{tabular}{@{}p{1.4cm}|c|c|cccc@{}}
\toprule
{}
& \multicolumn{1}{c|}{\textbf{MNIST}} 
& \multicolumn{1}{c|}{\textbf{CIFAR10}} 
& \multicolumn{4}{c}{\textbf{MIND}} \\
\cmidrule(lr){2-2} \cmidrule(lr){3-3} \cmidrule(lr){4-7}
& ACC & ACC & AUC & MRR & nDCG@5 & nDCG@10 \\
\midrule
FedSGD   & 99.38 & 82.03 & 65.20 & 30.72 & 33.58 & 40.09 \\
\cline{1-7}
Krum     & 95.29 & 73.46 & 52.91 & 23.12 & 24.23 & 30.56 \\
Median   & 97.09 & 62.13 & 54.05 & 23.31 & 24.36 & 31.17 \\
TMean    & 98.55 & 81.19 & 61.30 & 28.71 & 30.87 & 37.37 \\
FLTrust  & 98.08 & 80.56 & 62.66 & 29.13 & 31.66 & 38.07 \\
DPFLA    & 98.58 & 81.52 & 65.01 & 30.16 & 32.99 & 39.64 \\
FLGuardian & 98.09 & 81.40 & 64.61 & 30.70 & 33.49 & 39.91 \\
EnCAgg   & \textbf{98.60} & \textbf{81.57} & \textbf{65.17} & \textbf{30.75} & \textbf{33.61} & \textbf{40.15} \\
\bottomrule
\end{tabular}
}
\end{table}

\subsection{Experimental Results}
\subsubsection{Fidelity Evaluation}
To demonstrate the fidelity of our method, we first evaluate its performance alongside baseline methods under non-adversarial conditions. As shown in Table \ref{tab:fidelity}, our method achieves similar results to the baselines across different datasets.

The results indicate that the performance gap between our method and FedSGD is within 0.8 on image classification tasks, and within 0.3 on all metrics in the news recommendation task. Moreover, our method consistently outperforms all baselines across all evaluation metrics, demonstrating that it preserves comparable performance to standard federated learning algorithm FedSGD in non-adversarial settings and achieves strong fidelity.

\begin{table*}[htbp]
\centering
\caption{Robustness Evaluation on MNIST and CIFAR10 ( Accuracy (\%) )}
\label{tab:mnist_cifar_robust}
\renewcommand{\arraystretch}{1.2}
\setlength{\tabcolsep}{4pt}
\begin{tabular}{l|ccc|ccc|ccc|ccc}
\toprule
\rowcolor{white}
& \multicolumn{6}{|c|}{\textbf{10\% Malicious Clients}} 
& \multicolumn{6}{|c}{\textbf{60\% Malicious Clients}} \\
\cmidrule(lr){2-7} \cmidrule(lr){8-13}
\rowcolor{white}
& \multicolumn{3}{c|}{MNIST} 
& \multicolumn{3}{c|}{CIFAR10} 
& \multicolumn{3}{c|}{MNIST} 
& \multicolumn{3}{c}{CIFAR10} \\
\cmidrule(lr){2-4} \cmidrule(lr){5-7} \cmidrule(lr){8-10} \cmidrule(lr){11-13}
\rowcolor{white}
& UA-FedRec & LIE & Min-Max 
& UA-FedRec & LIE & Min-Max 
& UA-FedRec & LIE & Min-Max 
& UA-FedRec & LIE & Min-Max \\
\midrule
\rowcolor{white}
FedSGD   & 92.48 & 97.42 & 97.14 & 40.12 & 37.07 & 43.29 & 12.65 & 88.79 & 87.55 & 5.73  & 16.84 & 15.60 \\
\arrayrulecolor{black}
\cline{1-13}
\rowcolor{gray!10}
Krum     & 95.86 & 96.27 & 96.32 & 73.78 & 73.25 & 71.36 & 12.86 & 68.09 & 71.87 & 11.20 & 17.16 & 17.09 \\
\rowcolor{white}
Median   & 97.00 & 96.03 & 96.15 & 46.70 & 31.89 & 32.63 & 10.86 & 71.97 & 76.72 & 9.80 & 16.46 & 17.15 \\
\rowcolor{gray!10}
TMean    & 93.82 & 97.48 & 96.34 & 51.76 & 36.19 & 38.19 & 10.15 & 82.29 & 88.35 & 10.62 & 16.65 & 17.09 \\
\rowcolor{white}
FLTrust  & 98.06 & 97.24 & 97.29 & 80.09 & 56.48 & 53.44 & 97.49 & 91.80 & 89.94 & 79.44 & 20.19 & 18.95 \\
\rowcolor{gray!10}
DPFLA    & 98.48 & 97.44 & 97.22 & 81.40 & 39.53 & 41.38 & 12.81 & 77.55 & 72.60 & 2.59 & 17.07 & 16.87 \\
\rowcolor{white}
FLGuardian & 98.04 & 97.58 & \textbf{98.22} & 80.02 & 80.38 & 80.12 & 11.46 & 72.41 & 76.74 & 9.12 & 16.56 & 16.58 \\
\rowcolor{cyan!10}
EnCAgg & \textbf{98.58} & \textbf{98.23} & 98.16
& \textbf{81.86} & \textbf{81.01} & \textbf{80.58} 
& \textbf{98.32} & \textbf{97.91} & \textbf{97.78} 
& \textbf{81.54} & \textbf{80.69} & \textbf{80.39} \\
\bottomrule
\end{tabular}
\end{table*}

\begin{table*}[htbp]
\centering
\caption{Robustness Evaluation on MIND (\%)}
\label{tab:mind_robust}
\renewcommand{\arraystretch}{1.2}
\setlength{\tabcolsep}{3pt}
\rowcolors{4}{gray!10}{white}  
\resizebox{\linewidth}{!}{%
\begin{tabular}{l|cccc|cccc|cccc|cccc|cccc|cccc}
\toprule
\rowcolor{white}
& \multicolumn{12}{c|}{\textbf{10\% Malicious Clients}} 
& \multicolumn{12}{c}{\textbf{60\% Malicious Clients}} \\
\cmidrule(lr){2-13} \cmidrule(lr){14-25}
\rowcolor{white}
& \multicolumn{4}{c|}{UA-FedRec} & \multicolumn{4}{c|}{LIE} & \multicolumn{4}{c|}{Min-Max}
& \multicolumn{4}{c|}{UA-FedRec} & \multicolumn{4}{c|}{LIE} & \multicolumn{4}{c}{Min-Max} \\
\cmidrule(lr){2-5} \cmidrule(lr){6-9} \cmidrule(lr){10-13}
\cmidrule(lr){14-17} \cmidrule(lr){18-21} \cmidrule(lr){22-25}
\rowcolor{white}
& AUC & MRR & n@5 & n@10 & AUC & MRR & n@5 & n@10 & AUC & MRR & n@5 & n@10
& AUC & MRR & n@5 & n@10 & AUC & MRR & n@5 & n@10 & AUC & MRR & n@5 & n@10 \\
\midrule
\rowcolor{white}
FedSGD & 48.93 & 21.76 & 21.65 & 27.88 & 51.49 & 22.90 & 23.79 & 30.13 & 51.45 & 22.72 & 23.79 & 30.09 & 47.39 & 21.09 & 20.57 & 27.33 & 48.27 & 21.60 & 21.52 & 27.81 & 48.89 & 21.70 & 21.64 & 28.09 \\
\arrayrulecolor{black}
\cline{1-25}
\rowcolor{gray!10}
Krum & 58.38 & 25.39 & 27.49 & 34.14 & 51.81 & 24.14 & 24.97 & 31.01 & 51.94 & 22.22 & 23.03 & 29.58 & 47.61 & 20.28 & 19.46 & 26.48 & 49.13 & 21.77 & 21.89 & 28.09 & 48.78 & 21.63 & 21.59 & 27.95 \\
\rowcolor{white}
Median & 51.41 & 21.61 & 22.40 & 29.03 & 52.69 & 23.39 & 24.58 & 30.97 & 51.83 & 24.00 & 25.21 & 31.40 & 47.57 & 20.26 & 19.42 & 26.48 & 48.26 & 21.50 & 21.49 & 27.71 & 48.54 & 21.53 & 21.44 & 27.82 \\
\rowcolor{gray!10}
TMean & 50.46 & 23.14 & 23.66 & 29.53 & 54.16 & 23.22 & 24.62 & 31.27 & 50.67 & 22.40 & 23.40 & 29.63 & 47.47 & 21.11 & 20.61 & 27.37 & 48.44 & 21.57 & 21.52 & 27.89 & 49.00 & 21.70 & 21.69 & 28.13 \\
\rowcolor{white}
FLTrust & 55.42 & 22.55 & 23.73 & 30.88 & 60.20 & 26.98 & 29.13 & 35.79 & 55.03 & 23.93 & 24.86 & 31.52 & 53.24 & 23.10 & 24.00 & 30.80 & 53.09 & 22.71 & 23.80 & 30.49 & 52.49 & 22.60 & 23.68 & 30.17 \\
\rowcolor{gray!10}
DPFLA & 64.58 & 30.39 & 33.27 & 39.73 & 64.79 & 30.39 & 33.21 & 39.81 & 60.47 & 26.83 & 29.08 & 35.60 & 46.06 & 20.39 & 19.69 & 26.59 & 49.13 & 21.77 & 21.89 & 28.09 & 49.23 & 21.79 & 21.83 & 28.23 \\
\rowcolor{white}
FLGuardian & 63.66 & 29.52 & 32.21 & 38.72 & 60.98 & 27.53 & 29.76 & 36.55 & 60.99 & 27.44 & 29.49 & 36.12 & 47.60 & 20.43 & 19.79 & 26.68 & 47.92 & 21.19 & 21.15 & 27.37 & 48.67 & 21.56 & 21.52 & 27.89 \\
\rowcolor{cyan!10}
EnCAgg & \textbf{65.07} & \textbf{30.59} & \textbf{33.54} & \textbf{40.02} & \textbf{65.12} & \textbf{30.85} & \textbf{33.89} & \textbf{40.28} & \textbf{64.81} & \textbf{30.56} & \textbf{33.58} & \textbf{39.98} & \textbf{63.08} & \textbf{28.86} & \textbf{31.47} & \textbf{38.01} & \textbf{62.87} & \textbf{29.11} & \textbf{31.55} & \textbf{38.11} & \textbf{64.00} & \textbf{30.15} & \textbf{32.89} & \textbf{39.36} \\
\bottomrule
\end{tabular}}
\end{table*}

\subsubsection{Robustness Evaluation}
Table \ref{tab:mnist_cifar_robust} and Table \ref{tab:mind_robust} report the performance of all methods on MNIST, CIFAR-10, and MIND. We define the malicious client ratio as the fraction of clients capable of launching global poisoning attacks. These clients may randomly submit benign or malicious gradients in each round, so the ratio indicates the maximum possible proportion of malicious gradients.

Under low malicious ratios (10\%), many defenses fail to resist low-intensity attacks like LIE and Min-Max, which craft subtle perturbations to evade detection yet degrade the global model. For example, on CIFAR-10 under 10\% LIE and Min-Max, all methods except Krum and FLGuardian fall below 70.00 accuracy. However, EnCAgg overcomes this by blocking such perturbations from entering the benign cluster through carefully designed clustering and maintaining robustness. It achieves 81.01 and 80.58 under 10\% LIE and Min-Max on CIFAR-10 and outperforms all baselines on three datasets.

Under a high malicious ratio (60\%), most defenses break down as their benign-dominant assumptions no longer hold, and performance collapses. For example, on CIFAR-10 with 60\% malicious clients, almost all accuracies of other methods drops below 20 except FLTrust. In contrast, EnCAgg leverages known benign gradients to guide clustering without relying on majority assumptions, enabling stable defense against camouflaged and dynamic malicious strategies.As a result, it can maintain around 98 accuracy on MNIST and higher than 80 on CIFAR10, with only 1.0 drop from the 10\% setting on MIND, showing remarkable resilience.

\begin{figure}
  \centering
  \includegraphics[width=1.0\linewidth]{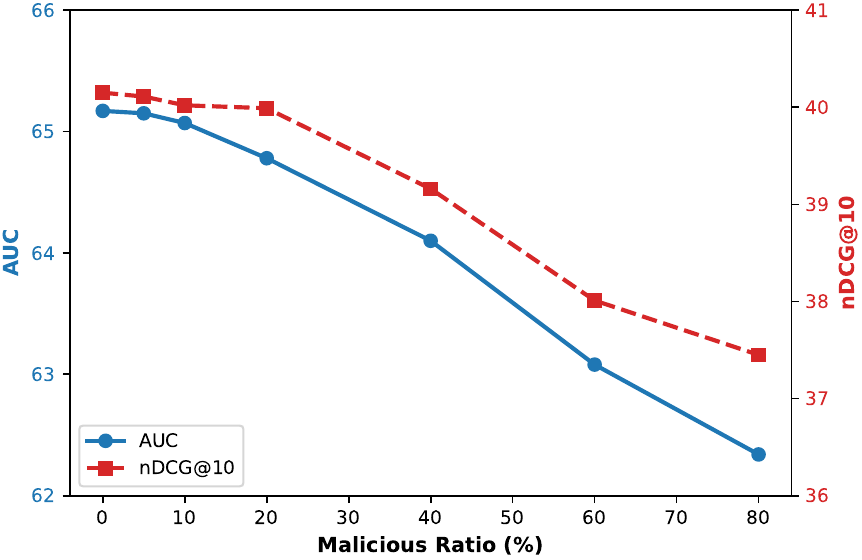}
  \caption{Impact of malicious client ratio on EnCAgg}
  \label{fig:mal_ratio}
\end{figure}

\subsubsection{Impact of Malicious Client Ratio}
To further evaluate robustness under varying malicious client ratios $mal$, we test our method on MIND under UA-FedRec attacks with 5\%, 10\%, 20\%, 40\%, 60\%, and 80\% malicious clients.

As shown in Fig. \ref{fig:mal_ratio}, our method remains consistently robust across the entire range. At 5\% malicious clients, metrics reach AUC = 65.25, MRR = 30.95, nDCG@5 = 33.99, and nDCG@10 = 40.41, which almost matching no attack scenario. Even at 80\% malicious clients, they remain as high as 62.34, 28.50, 30.88, and 37.45, with no drop exceeding 3.0. This degradation stems from reduced effective benign participation, which limits usable training data and, due to heterogeneity, causes global updates to deviate from the true data distribution. Still, the performance drop remains mild, preserving overall model utility. These results confirm our method’s strong robustness, even under extreme attack scenarios with up to 80\% malicious clients.

\subsubsection{Impact of Radius Coefficient}
In our clustering mechanism, the radius coefficient $r$ is a key hyperparameter that determines the clustering radius $\varepsilon$ by selecting two known benign gradients. To assess its impact under different malicious ratios, we conduct experiments on CIFAR-10 and MNIST using LIE and Min-Max attacks at 10\% and 60\% malicious client participation. For each setting, $r$ ranges from 0.1 to 0.9 (step size: 0.1), and classification accuracy is recorded.

\begin{figure}
  \centering
  \includegraphics[width=1.0\linewidth]{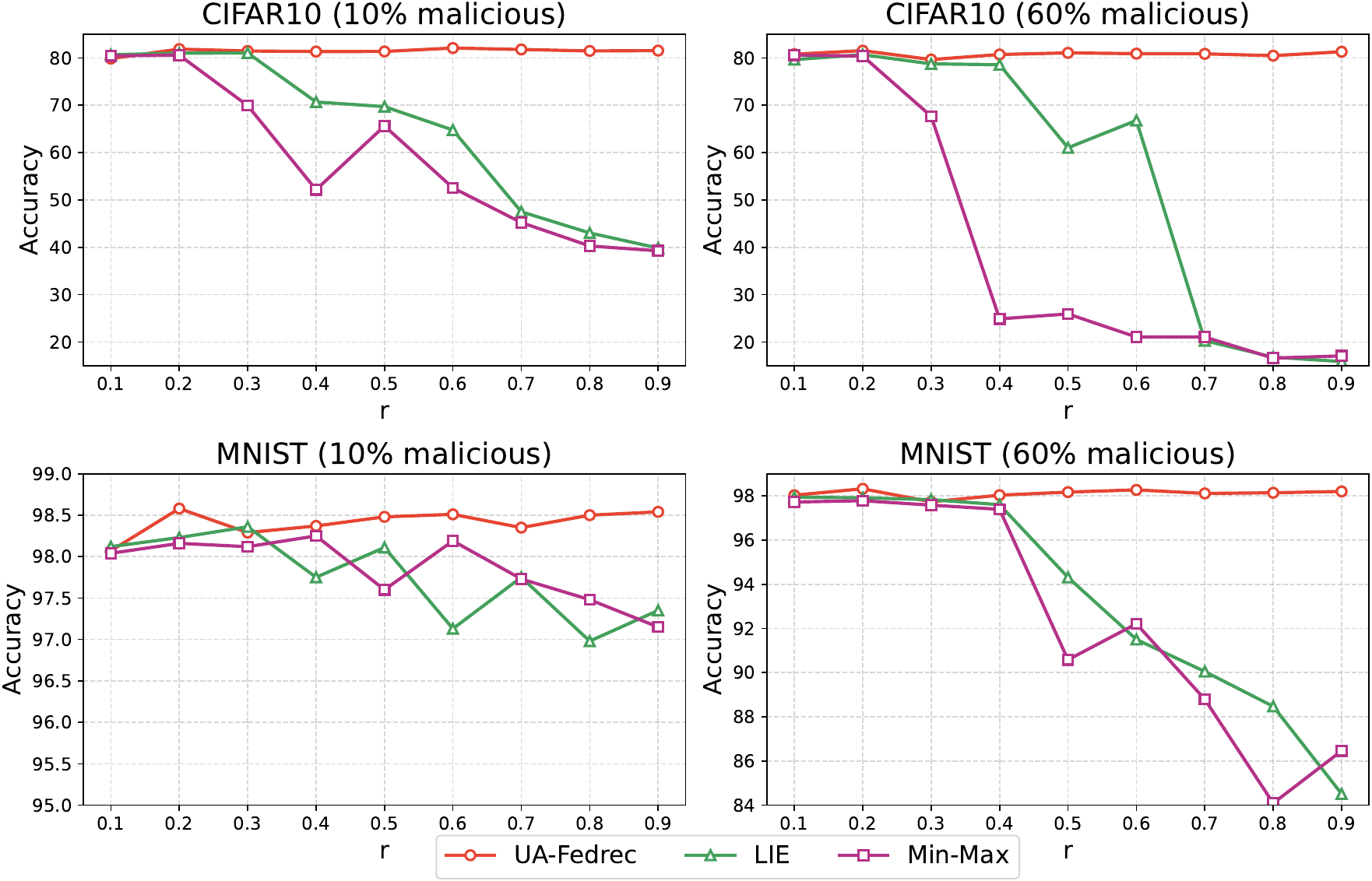}
  \caption{Impact of malicious client ratio on our method}
  \label{fig:r_value}
\end{figure}

As shown in Fig. \ref{fig:r_value}, accuracy remains stable when $r$ is moderately small (0.1–0.3), regardless of dataset or malicious ratio. This aligns with our design: a smaller $r$ yields a tighter $\varepsilon$, excluding stealthy malicious gradients while still preserving most tightly packed benign ones. In contrast, larger $r$ degrades performance under stealthy attacks, especially at high malicious ratios. This is because such attacks craft perturbations close to sparse benign updates. A larger $r$ leads to a larger $\varepsilon$, which increases the risk of adversarial gradients being clustered with benign ones and degrades global model performance. This poisoning effect becomes more severe as the malicious client ratio rises. With an appropriate $r$, EnCAgg effectively resists various poisoning attacks across different malicious ratios and attack intensities.

\subsubsection{Adaptive Attack}

As EnCAgg mainly relies on low-dimensional clustering results to filter gradients, adversaries can construct adaptive attacks specifically targeting this aggregation rule. In this work, we adopt the AGR Tailored Attack idea proposed by Shejwalkar \textit{et al.}~\cite{shejwalkar2021manipulating}, where the attacker is assumed to have full knowledge of the server-side aggregation rule and benign clients' gradients, and exploits this information to craft malicious gradient accordingly. In this setting, the attacker can not only access all or part of the gradients submitted by other clients, but also obtain the aggregation rule of EnCAgg. Based on this information, the attacker locally simulates the dimensionality reduction, clustering, and filtering process of EnCAgg, and further adjusts malicious gradients to increase their probability of passing the clustering filter and participating in the final aggregation.

The proposed adaptive attack mainly exploits the low-dimensional clustering and high-dimensional aggregation characteristic of EnCAgg. Specifically, the attacker first estimates the benign cluster distribution in the projected two-dimensional space according to benign gradients and selects benign reference gradient. The malicious gradients are then constructed to remain close to benign gradients in the low-dimensional projection space so as to satisfy the clustering radius and density constraints, while simultaneously deviating in the original high-dimensional space to influence the final aggregation result. Specifically, the adversarial update is constructed as
\begin{equation}
\mathbf{g}_{adv}
=
\mathbf{g}_{ref}
+
\delta_{\parallel}
+
\delta_{\perp},
\end{equation}
where $\mathbf{g}_{ref}$ denotes the benign reference gradient, $\delta_{\parallel}$ represents the low-dimensional perturbation within the PCA principal subspace, which is used to adjust the position of malicious updates in the two-dimensional projection space so that they remain close to the benign cluster and pass the clustering filter, and $\delta_{\perp}$ denotes the high-dimensional perturbation in the orthogonal complement space of the PCA principal subspace. This perturbation is used to interfere with the final high-dimensional aggregation result while maintaining low-dimensional similarity to benign gradients, thereby enhancing the attack effectiveness. To prevent the high-dimensional perturbation from affecting the clustering result in the projected space, the attacker further constrains
\begin{equation}
\mathbf{V}_2^\top
\delta_{\perp}
\approx 0.
\end{equation}
As a result, $\delta_{\perp}$ remains nearly invisible in the two-dimensional projection space while still influencing the final high-dimensional aggregation process. The perturbation strength is further adaptively adjusted according to the locally simulated clustering results to balance attack effectiveness and stealthiness.

Based on the above adaptive attack setting, we evaluate the robustness of different aggregation methods. Experimental results are summarized in Table~\ref{tab:adaptive_attack}. Under a $10\%$ malicious client ratio, all methods maintain relatively high accuracy, while EnCAgg achieves the best performance with $97.51\%$ accuracy. When the malicious ratio increases to $60\%$, most methods degrade significantly, whereas EnCAgg still maintains a high accuracy of $94.67\%$, demonstrating strong robustness against adaptive attacks. These results indicate a trade-off between attack effectiveness and low-dimensional stealthiness, which limits the ability of adaptive attacks to fully bypass EnCAgg.

\begin{table}[htbp]
\centering
\caption{Performance under Adaptive Attack (\%)}
\label{tab:adaptive_attack}
\begin{tabular}{lcc}
\toprule
\textbf{Method} & \textbf{10\% Malicious} & \textbf{60\% Malicious} \\
\midrule
FedSGD        & 95.47 & 73.11 \\
Krum          & 94.65 & 69.71 \\
Median        & 95.14 & 70.99 \\
Trimmed Mean  & 96.13 & 73.97 \\
FLTrust       & 95.94 & 76.08 \\
DPFLA         & 97.20 & 70.80 \\
FLGuardian    & 97.37 & 71.93 \\
EnC           & \textbf{97.51} & \textbf{94.67} \\
\bottomrule
\end{tabular}
\end{table}

\subsubsection{Ablation Study}

\begin{figure}
  \centering
  \includegraphics[width=0.8\linewidth]{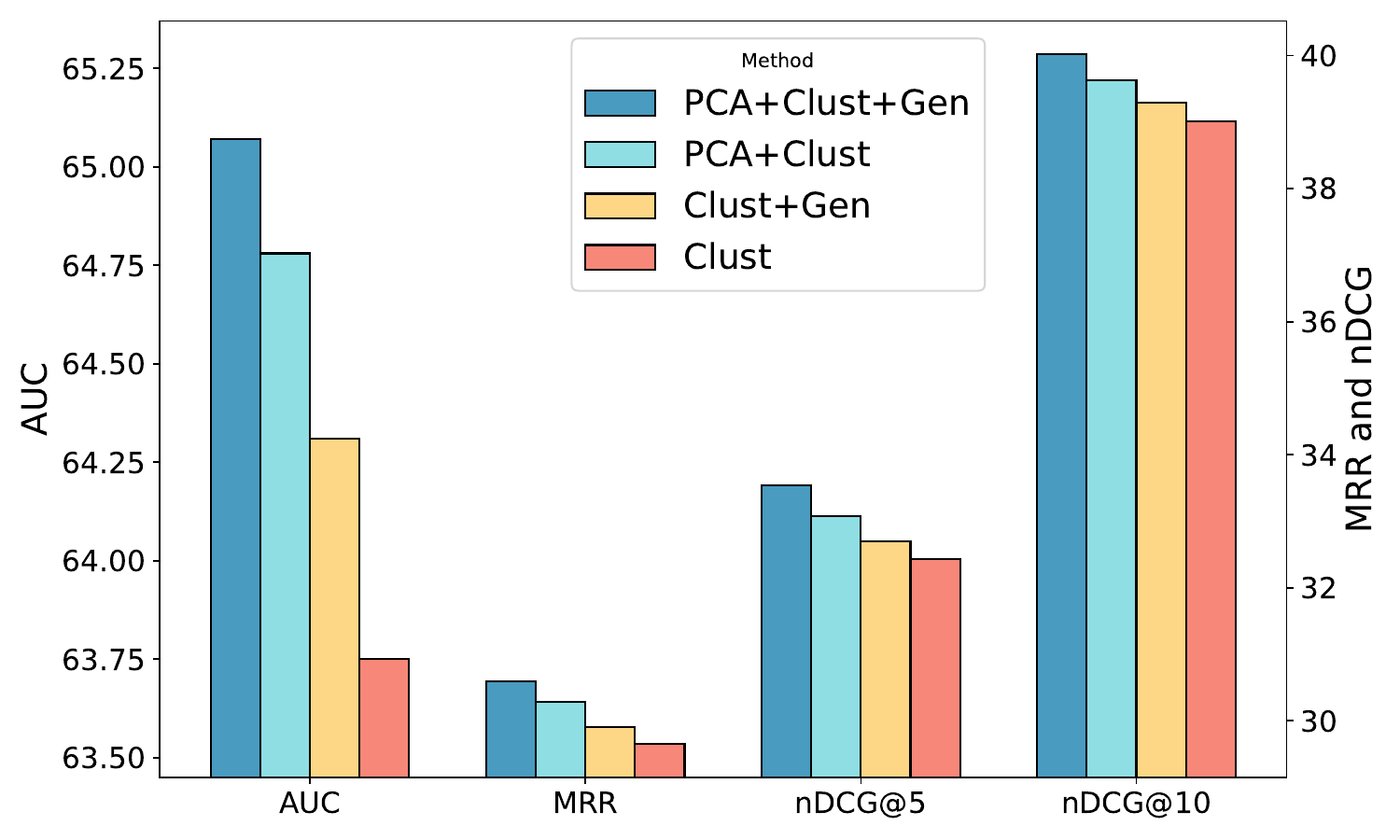}
  \caption{Ablation study}
  \label{fig:ablation}
\end{figure}

Finally, we conduct ablation studies on the MIND dataset with 10\% malicious clients performing UA-FedRec attacks to evaluate the contributions of dimensionality reduction and the gradient generator. As shown in Fig. \ref{fig:ablation}, removing the generator causes an approximately 0.1 performance drop. This is because pseudo-gradients help pull sparsely distributed benign updates into the benign cluster during the second round of clustering. Without them, these updates are treated as noise and excluded from aggregation, resulting in incomplete global updates.

Removing dimensionality reduction causes a larger performance drop of approximately 0.7. Dimensionality reduction improves the separability between benign and malicious gradients. Without it, attackers can more easily craft gradients within the clustering radius, bypassing the filtering mechanism and degrading model performance.

When both components are removed, the performance drops by approximately 1.3, indicating that density-based clustering alone is insufficient for robust defense. The results demonstrate that dimensionality reduction and pseudo-gradient generation jointly contribute to the effectiveness of EnCAgg.

\section{Conclusion}
In this paper, we proposed EnCAgg, a novel robust aggregation framework for federated learning, which effectively defends against dynamic model poisoning attacks. EnCAgg consists of low-dimensional gradient clustering and pseudo-gradient generation to accurately distinguish and retain benign updates, while filtering out a wide range of adversarial ones. Extensive experimental results on multiple datasets under various attack and parameter settings demonstrate that EnCAgg outperforms existing baselines in both fidelity and robustness against novel model poisoning strategies, especially under high proportions of malicious clients. For future work, we plan to extend the generator model’s capability and diversity against multiple types of sophisticated attacks.

\bibliographystyle{IEEEtran}
\bibliography{Biblio-Bibtex}

\vfill

\end{document}